\documentclass{PoS} 
\usepackage{wrapfig} 
\title{
{\small{DESY 19--152, DO-TH 19/16, SAGEX-19-21}} \\
An Update of the ABM16 PDF Fit}

\ShortTitle{An Update of the ABMP16 PDF Fit}

\author{\speaker{Sergey Alekhin}%
         \thanks{This work was supported in part by Bundesministerium f\"ur Bildung und 
                 Forschung (contract 05H18GUCC1), by EU TMR network SAGEX
                 agreement No. 764850 (Marie Sk\l{}odowska-Curie) and COST action CA16201: Unraveling new
physics at the LHC through the precision frontier.
}\\ 
II. Institut f\"ur Theoretische Physik, Universit\"at Hamburg,
    Luruper Chaussee 149, D-22761 Hamburg, Germany;\\
        Institute for High Energy Physics,142281 Protvino, Russia\\
        E-mail: \email{sergey.alekhin@desy.de}}

\author{Johannes Bl\"umlein\\ 
        Deutsches Elektronensynchrotron DESY, Platanenallee 6, D--15738 Zeuthen, Germany\\
        E-mail: \email{Johannes.Bluemlein@desy.de}}

\author{Sven-Olaf Moch\\
II. Institut f\"ur Theoretische Physik, Universit\"at Hamburg,
    Luruper Chaussee 149, D-22761 Hamburg, Germany \\      
 E-mail: \email{sven-olaf.moch@desy.de}}

\abstract{We present an updated version of the ABMP16 nucleon PDFs, which is tuned by using recent 
precise data on $W$- and $Z/\gamma^*$-production at the LHC and the final HERA data on  DIS 
$c$- and $b$-quark production and by imposing a stringent $Q^2$-cut on the inclusive DIS 
data in order to avoid the impact of higher twist terms at small $x$ at HERA. The new 
$W$- and $Z$-boson production data, in particular the updated version of the ATLAS data at 
the c.m.s. energy 7 TeV, are well accommodated into the present fit. The strange sea 
distribution obtained is consistent with the average of the up and down quark ones at small 
$x$. However, it is still suppressed with respect to the non-strange one by a factor of $\sim 0.5$ 
at moderate $x$. The small-$x$ gluon distribution is enhanced as compared to the previous 
ABMP16 fit, in line with updated data on the DIS $c$-quark production. Finally, a good 
description of the non-resonant $\gamma^*/Z$-production data, which are included into the 
ABM analysis for the first time, is achieved provided the photon-initiated lepton pair 
production is taken into account.
}

\FullConference{XXVII International Workshop on Deep-Inelastic Scattering and Related Subjects (DIS2018)\\
		08-12 April 2019\\
		Torino, Italy}

\begin{document}

The ABMP16 proton parton distribution functions (PDFs)~\cite{Alekhin:2017kpj} are extracted from a 
combination of data on inclusive neutral-current (NC) deep-inelastic scattering (DIS), semi-inclusive $c$- 
and $b$-quark for NC and charged-current (CC) DIS data and $W$-, $Z$-boson and single/double 
$t$-quark production in (anti)proton-proton collisions. The data set used has 
a high accuracy of typically ${\cal O}$(1\%) and allows the precise determination of the quark distributions in 
a wide range of parton momentum fractions $x$ and of the gluon distribution at small and moderate $x$. Nonetheless, 
the analysis can still be improved by using the advantage of the fast growing statistics of the LHC experiments. 
In this paper we describe such an update with the focus on the impact of recent $W$- and $Z/\gamma^*$-boson 
production data. We also check updated measurements of the $c$- and $b$-quark production at HERA and consider 
a new treatment of the power corrections (higher twist) to the DIS cross sections. 
\begin{figure}
  \centering
  \includegraphics[width=\textwidth,height=0.4\textwidth]{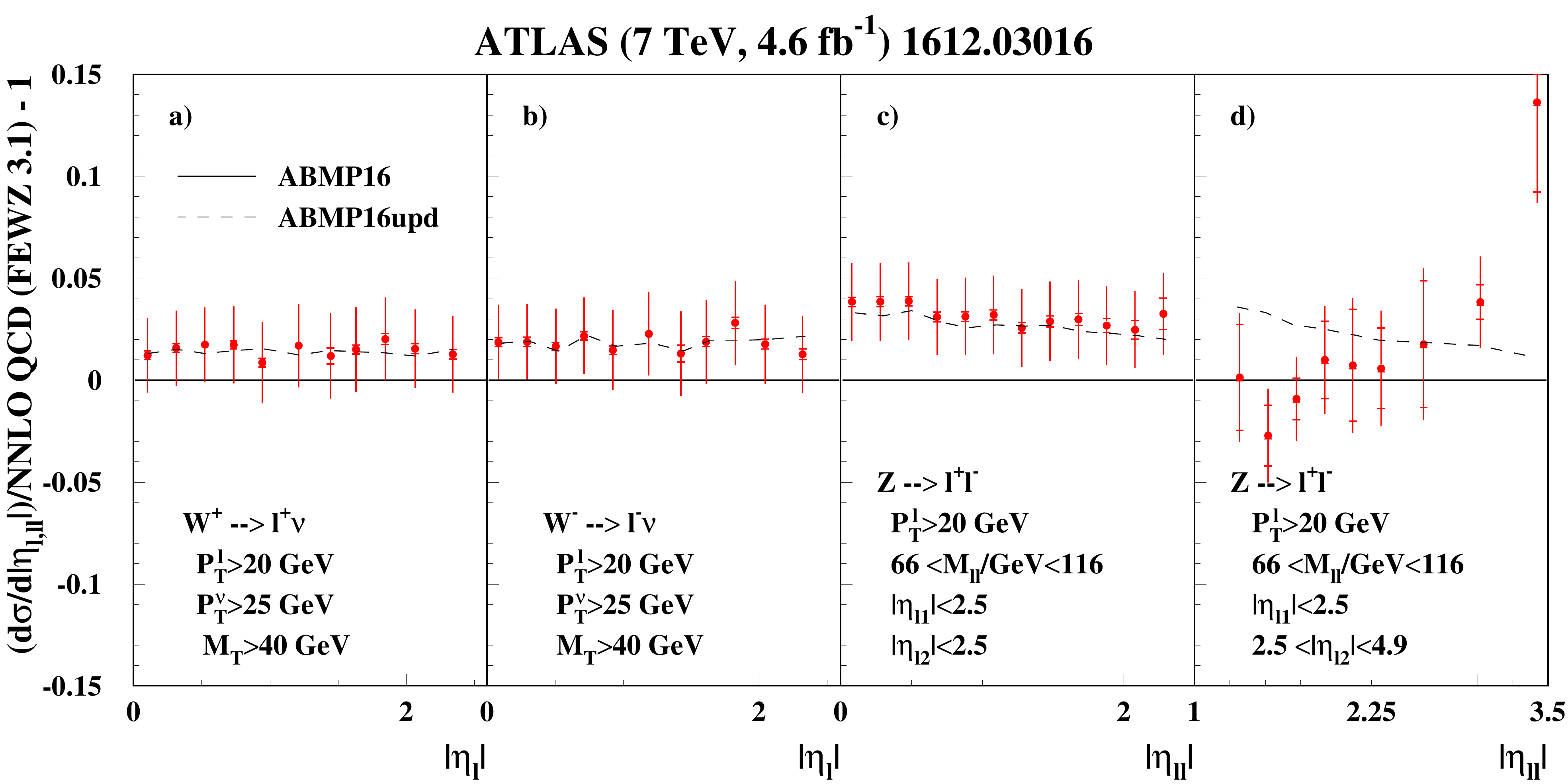}
  \caption{\small The pulls of the $W$- and 
$Z$-boson production ATLAS data of 
Ref.~\cite{Aaboud:2016btc} in various boson-rapidity regions
w.r.t. the NNLO QCD predictions obtained by using 
{\tt FEWZ~3.1} in combination with the ABMP16 
PDFs~\cite{Alekhin:2017kpj} (solid lines) and the ABMP16upd ones 
obtained in the present analysis (dashed) 
({\bf a)}: $W^+$ (central), {\bf b)}: $W^-$ (central), {\bf c)}: $Z$ (central), {\bf d)}:
$Z$ (forward).
}
    \label{fig:atlaswz}
\end{figure}

\begin{figure}
  \centering
  \includegraphics[width=\textwidth,height=0.35\textwidth]{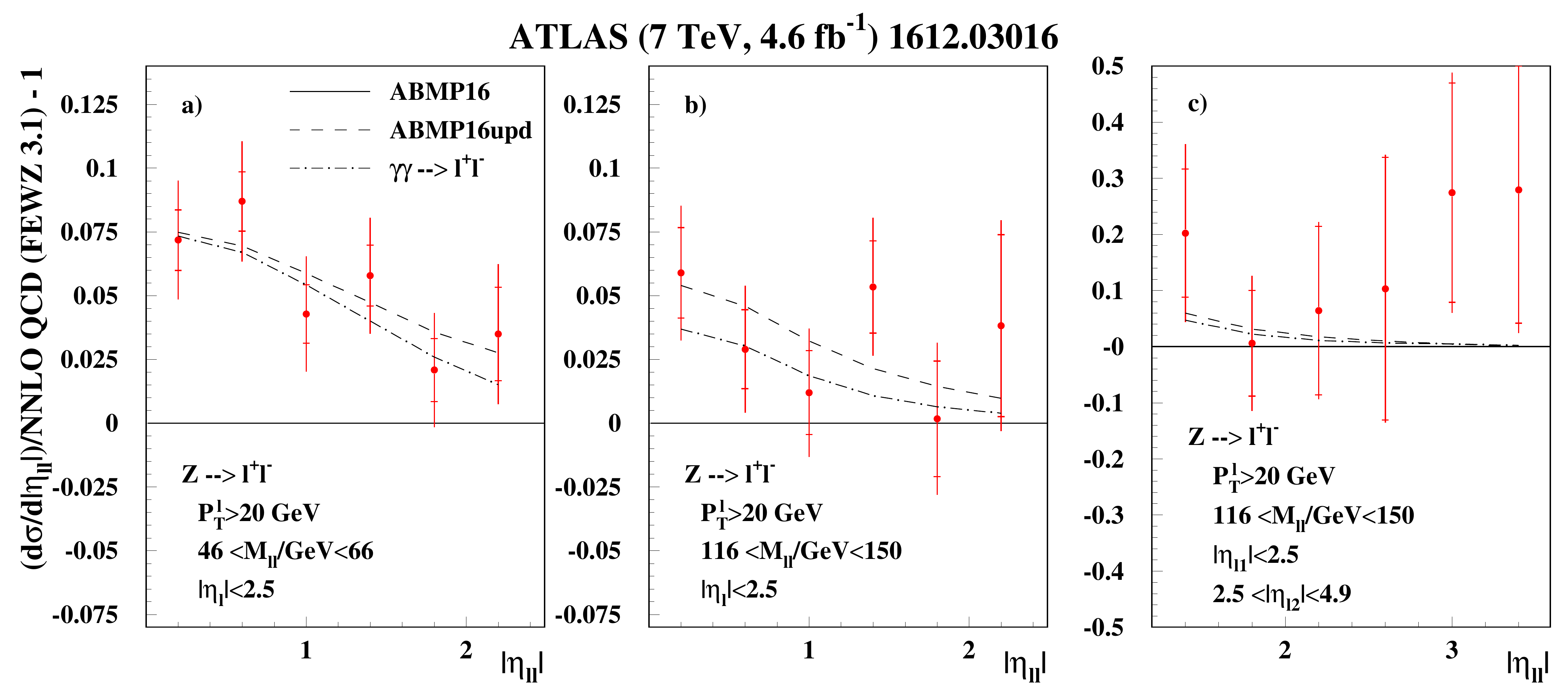}
  \caption{\small The same as Fig.~{\protect \ref{fig:atlaswz}} for 
the non-resonant DY data in various bins on the 
lepton pair invariant mass $M_{ll}$ and pseudo-rapidity  $\eta_{ll}$
({\bf a)}: $46<M_{ll}/{\rm GeV}<66$ (central), {\bf b)}: 
$116<M_{ll}/{\rm GeV}<150$ (central), {\bf c)}:
$116<M_{ll}/{\rm GeV}<150$ (forward)). 
The LO contribution of the photon-induced channel
$\gamma \gamma \rightarrow l^+ l^-$ computed with the photon distributions
obtained n the present analysis is given for comparison (dashed dots).
}
    \label{fig:offz}
\end{figure}

An important component of the present update are recent high-statistics Drell-Yan data collected at the LHC. 
In particular, the ATLAS collaboration greatly improved the accuracy of their 7~TeV $W$- and $Z$-boson 
sample~\cite{Aaboud:2016btc}, which now supersedes the earlier data set released in 2011~\cite{Aad:2011dm}. 
Besides, the data of Ref.~\cite{Aaboud:2016btc} cover the non-resonant $\gamma^*$ channel, which provides an 
additional independent constraint on the PDFs. The whole ATLAS data set is very well accommodated into the 
ABMP16 fit with the value of $\chi^2/NDP=69/61$ obtained at next-to-next-to-leading order (NNLO) QCD. The  
irreducible background from the process $\gamma \gamma \rightarrow l^+ l^-$, where $l^{\pm}$ denotes charged leptons,
significantly improves the agreement of the fit with the data, cf.~Fig.~\ref{fig:offz}. In the present analysis 
leading-order QED corrections are used and the photon distribution is included into the present PDF fit.
The photon distribution preferred by the non-resonant DY data within this approach are evidently larger than 
MRST2004qed ones~\cite{Martin:2004dh} employed to correct the ATLAS data~\cite{Aaboud:2016btc} for photonic 
initial-state contribution.
No statistically significant trends are seen in the pulls, cf. Figs.~\ref{fig:atlaswz},~\ref{fig:offz}, 
except of the forward $Z$-boson production case. Due to large uncertainties in the data the value of $\chi^2$ for 
this sample is still considered to be reasonable. 

\begin{figure}
  \centering
  \includegraphics[width=0.48\textwidth,height=0.45\textwidth]{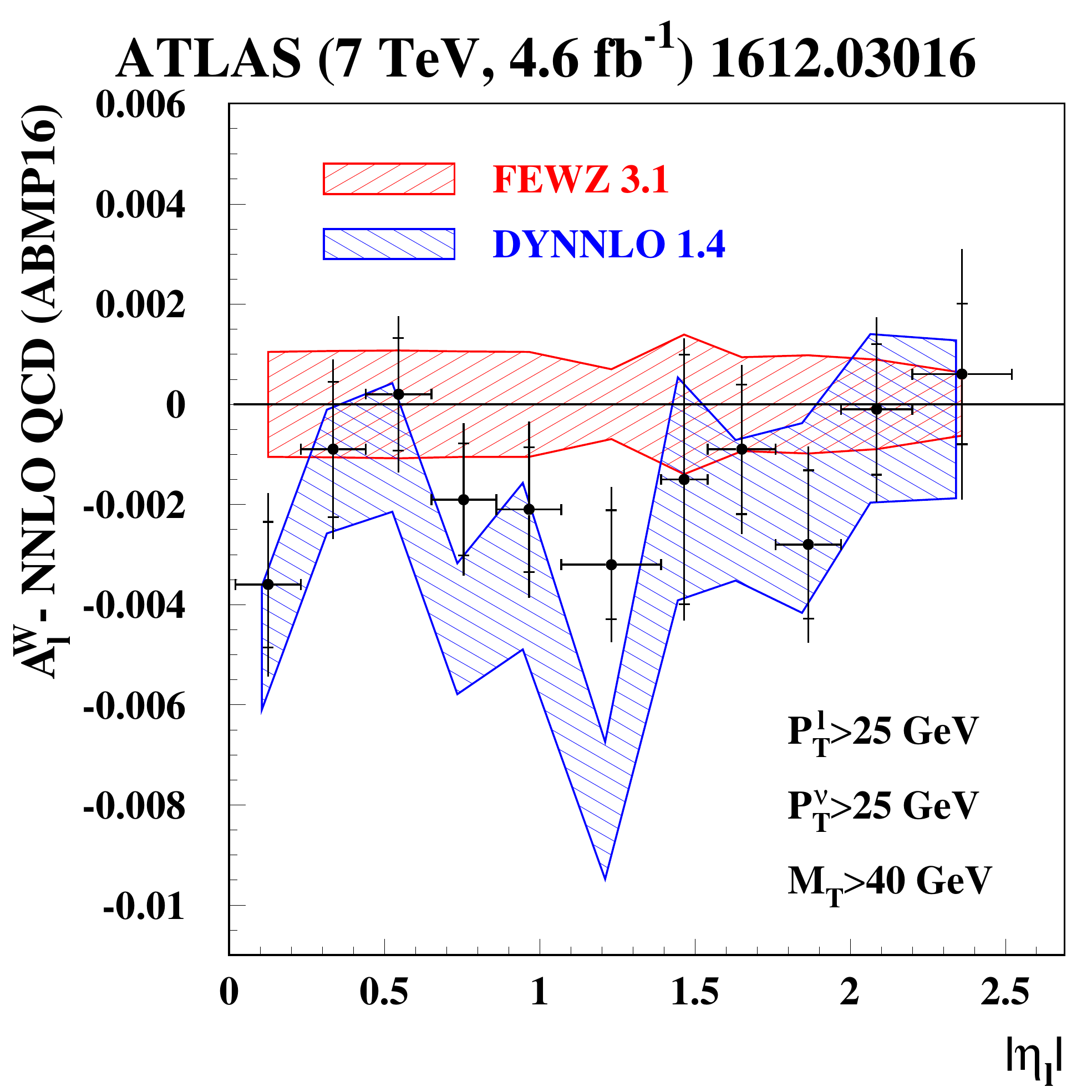}
  \includegraphics[width=0.47\textwidth,height=0.45\textwidth]{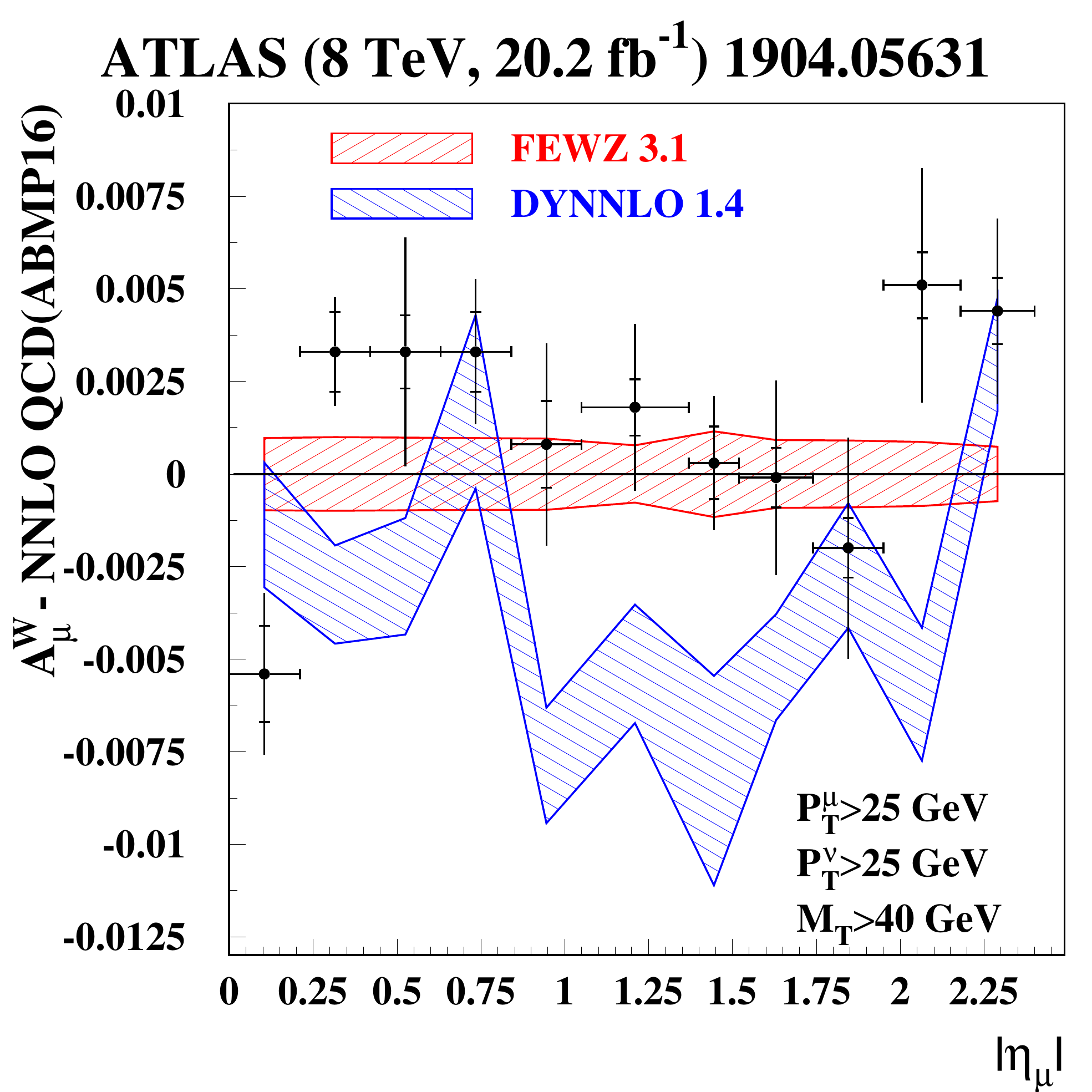}
  \caption{\small The same as Fig.~{\protect \ref{fig:atlaswz}} for 
the ATLAS data on the lepton asymmetry $A_l$ collected at c.m.s. energy 7 
TeV~\cite{Aaboud:2016btc} (left panel) and muon asymmetry $A_\mu$ 
collected at c.m.s. energy 8 TeV~\cite{Aad:2019rou} (right panel) and with 
various NNLO tools: {\tt FEWZ~3.1}~\cite{Li:2012wna} (right-tilted hatch) and 
{\tt DYNNLO~1.4}~\cite{Catani:2009sm} (left-tilted hatch). The bands in 
theory predictions represent computational uncertainties obtained 
with ${\cal O}$(month) computer wall time.
}
    \label{fig:asym78}
\end{figure}

It is worth mentioning that, given the experimental accuracy achieved by now, a comparison of the data with 
the QCD predictions is sensitive to the choice of the theoretical tool used. It is particularly relevant 
for the data on the charged lepton asymmetry $A_l$, where many systematic uncertainties cancel in the ratio. 
Indeed, a comparison of the NNLO QCD predictions with the ATLAS data on $A_l$ collected at 7 and 8~TeV collision 
energy~\cite{Aaboud:2016btc,Aad:2019rou} demonstrate different trends for two publicly available tools, 
{\tt FEWZ~3.1}~\cite{Li:2012wna} and {\tt DYNNLO~1.4}~\cite{Catani:2009sm}. The FEWZ predictions do somewhat 
overshoot the data at 7 TeV, while the DYNNLO ones go lower and are in better agreement with the measurements. 
At 8 TeV the tendency is different: The FEWZ predictions somewhat undershoot the data and the DYNNLO ones go 
essentially lower, cf. Fig.~\ref{fig:asym78}. In summary, the FEWZ predictions demonstrate a better overall 
agreement with the data. Therefore this tool is routinely used in our fit. 

\begin{figure}
  \centering
  \includegraphics[width=\textwidth,height=0.9\textwidth]{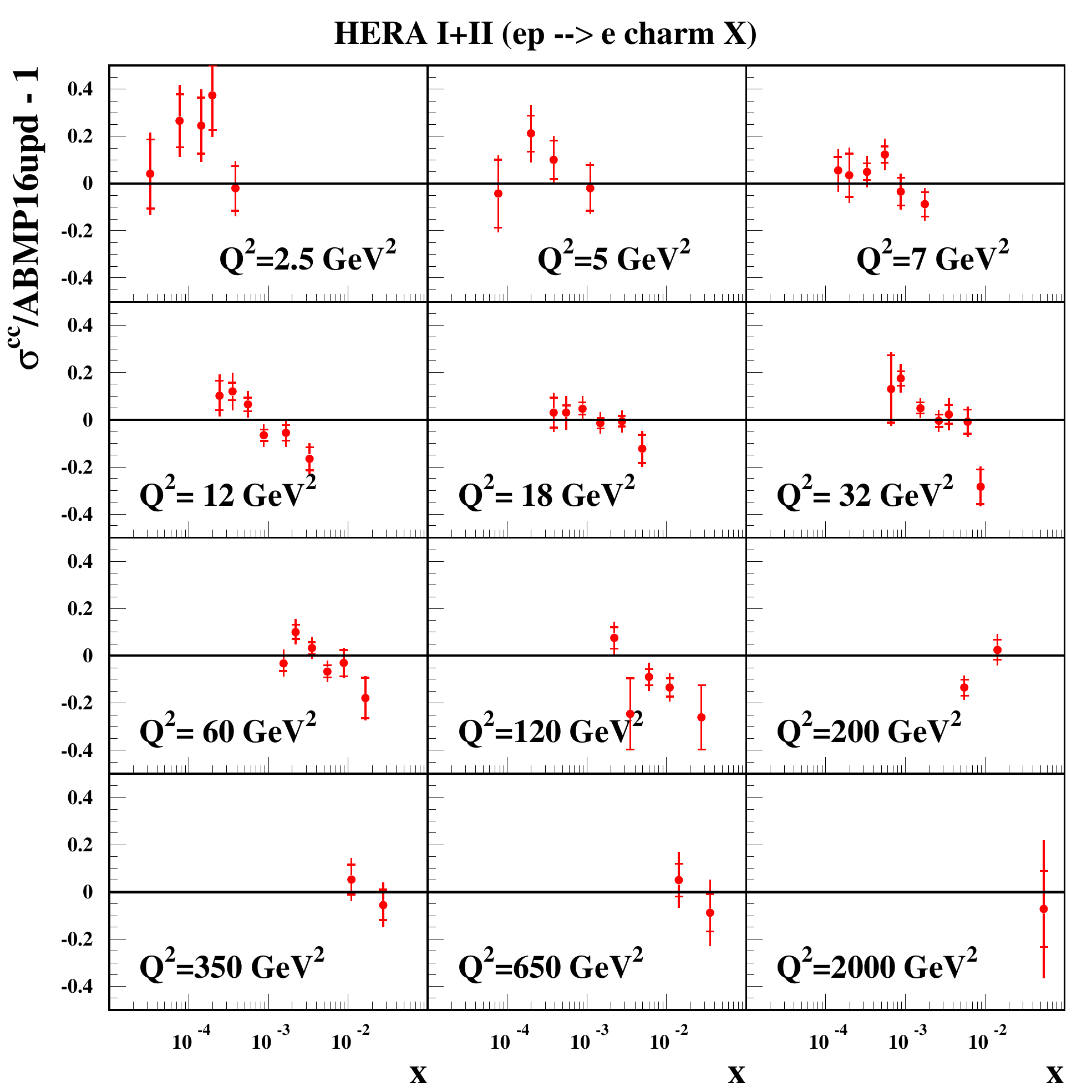}
  \caption{\small The pulls of the combined HERA data on the DIS 
$c$-quark production ~\cite{H1:2018flt} w.r.t. the ABMP16upd 
predictions obtained in the present fit versus Bjorken $x$ and 
for various bins in the virtualities $Q^2$ .
}
    \label{fig:plotcc}
\end{figure}

One more improvement is an update of the HERA data on semi-inclusive $c$- and $b$-quark DIS 
production~\cite{H1:2018flt}. This process is dominantly initiated by gluons. Therefore these data  impose an 
additional constraint on the gluon distribution at small $x$, which is otherwise predominantly driven by the slope
of the DIS inclusive structure function $F_2$ w.r.t. $\ln(Q^2)$. As we have shown earlier, the factorization scheme with 
three light quarks in the initial state provides a solid theoretical framework for the description of the DIS 
heavy-flavor production~\cite{HQ}. Furthermore, the value of the $c$-quark mass in the $\overline{\rm MS}$-scheme
extracted from the experimental data within this framework is in very good agreement with other determinations. 
This gives us additional confidence in the validity of this approach.
The HERA data on the DIS $c$-quark production~\cite{H1:2018flt} 
are also in a reasonable agreement with our updated fit with the value of  
$\chi^2/NDP=134/79$ obtained for the whole sample. The pulls of these 
data w.r.t. the present fit have no systematic trend and 
look rather like statistical fluctuations, which sometimes go 
beyond the uncertainties and therefore pull up the value of $\chi^2$.
For example, 
the slope in Bjorken $x$ observed for the bins with momentum transfer
$Q^2=12, 32~{\rm GeV}^2$ is not confirmed in the neighbor  bins
with $Q^2=7, 18~{\rm GeV}^2$, cf. Fig.~\ref{fig:plotcc}.

\begin{figure}
  \centering
  \includegraphics[width=0.48\textwidth,height=0.4\textwidth]{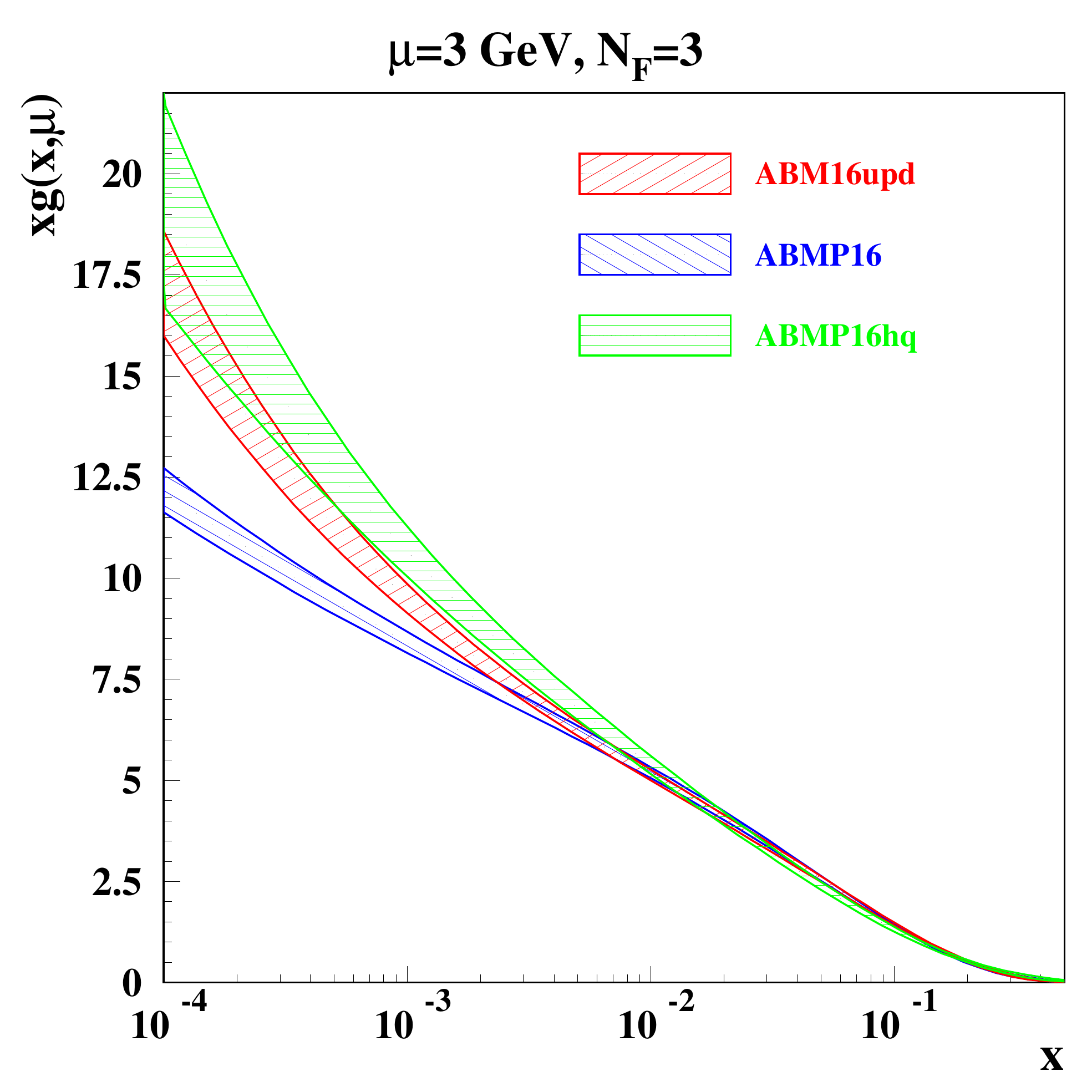}
  \includegraphics[width=0.47\textwidth,height=0.4\textwidth]{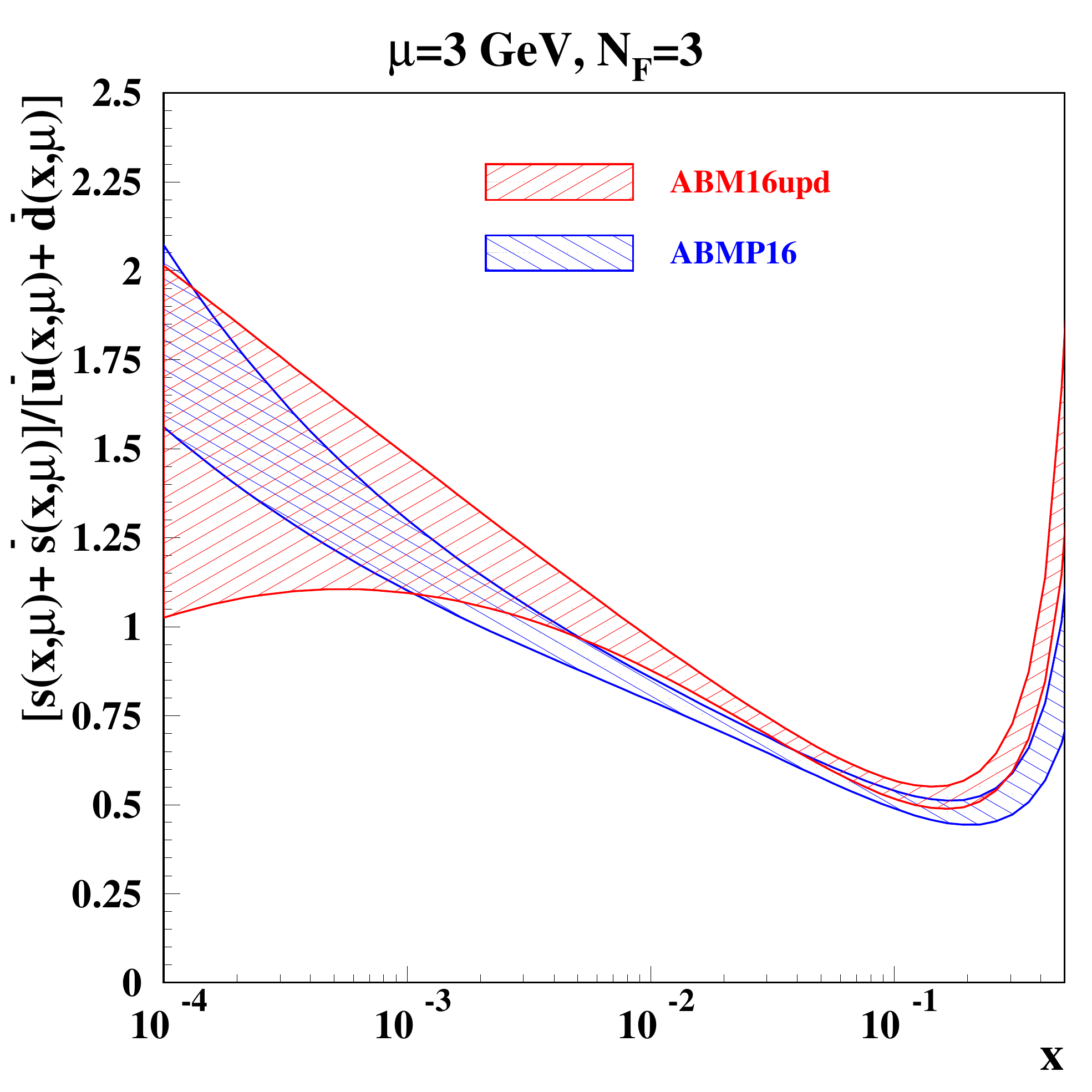}
  \caption{\small Left: The error band for the ABMP16upd 3-flavor 
gluon distribution versus $x$ 
at the factorization scale $\mu=3~{\rm GeV}$ obtained
in the present fit (right-tilted hatch) in comparison with the ABMP16
one (left-tilted hatch) and ABMP16hq one (horizontal hatch), 
which was obtained in the variant of present analysis with the HERA DIS 
inclusive data dropped. Right: The same as the left panel for the 
strange sea suppression factor 
$[s(x,\mu)+ \bar s(x,\mu)]/[\bar u(x,\mu)+ \bar d(x,\mu)]$, where 
$u(x,\mu)$, $d(x,\mu)$ and $s(x,\mu)$ are distributions of the 
up, down and strange quarks, respectively.
}
    \label{fig:pdfs}
\end{figure}

It is worth noting that the charm production data prefer 
a steeper small-$x$ gluon distribution as compared to the 
one obtained in the ABMP16 fit. A detailed examination shows that the tension 
is driven by the inclusive HERA data at small $x$ and $Q^2$. This kinematic 
region is potentially problematic w.r.t. perturbative QCD analyses in view of the 
relatively large value of strong coupling constant $\alpha_s$ and 
singularities in the 
anomalous dimensions and Wilson coefficients at small $x$.
A phenomenological description of this part of the inclusive data 
in the ABMP16 fit includes power corrections in form 
of a higher twist contribution parameterized in a model-independent form 
and fitted to the data simultaneously with the leading twist PDFs.
The twist-4 term preferred by the inclusive HERA data is 
negative at small $x$~\cite{Alekhin:2017kpj,Abt:2016vjh}, which  
demonstrates a trend of damping the $Q^2$-slope in $F_2$ driven by the small $x,Q^2$ part
of this sample. 
In the updated version of our fit a more stringent cut 
$Q^2>10~{\rm GeV}^2$ is imposed on the inclusive DIS data, cf. also \cite{Alekhin:2012ig}. This allows to 
neglect the higher twist contributions and
circumvents in such a way potential problems of small-$Q^2$ phenomena.
This approach leads also to a better agreement between inclusive and 
semi-inclusive HERA data sets. In particular, 
this results in the gluon distribution, which 
is consistent to the one extracted from the $c$- and $b$-quark 
production data without taking into account the inclusive ones, 
cf. Fig.~\ref{fig:pdfs}.

%
%

The quark distributions are in general not sensitive to this modification 
of the fit ansatz. However, the updated strange sea distribution  
is smaller than the one of  ABMP16 at small $x$ and consistent with the 
average of up- and down-quark distributions. Meanwhile, 
the strange sea is still suppressed w.r.t. the non-strange one
by factor of $\sim 0.5$ at moderate $x$  and  
this leads to overall strange-sea suppression by an integral factor 
$\kappa_s(\mu^2=20~{\rm GeV^2})=0.71(3)$, where
$$
\kappa_s(\mu^2) \, = \,\frac{\int_0^1 x[s(x,\mu^2) + \bar s(x,\mu^2)]dx}{\int_0^1 x[\bar d(x,\mu^2) + \bar u(x,\mu^2)]dx}
\, .
$$
This is in agreement with the ABMP16 result and our earlier study of the 
impact of the ATLAS data of Ref.~\cite{Alekhin:2017olj} 
on the strange sea determination. 
Note that the PDFs with such a suppression provide a much better description
of ATLAS data  as 
compared to the analysis of Ref.~\cite{Aaboud:2016btc} reporting the
value of $\chi^2/NDP=108/61$ for the PDF set, which implies 
an enhanced strange sea. 

In summary, we report on an updated version of the ABM PDF fit, which includes recent 
precise data on $W$- and $Z/\gamma^*$-production at the LHC and final 
HERA data on DIS $c$- and $b$-quark production. Also a stringent 
cut on $Q^2$ is imposed on the inclusive DIS data in order to 
avoid any impact of the higher twist terms at small $x$, contained in the HERA data.
The recent precise $W$- and $Z$-boson production data 
from the LHC, in particular the updated version of the ATLAS data at c.m.s. 
energy 7 TeV, are well accommodated into the present fit.
The latter prefer a strange sea quark distribution, which is 
consistent with the average of up and down ones at small $x$ and which is 
suppressed by a factor of $\sim 0.5$ at moderate $x$. The small-$x$ gluon distribution 
is enhanced as compared with the previous ABMP16 fit due to a more stringent cut 
imposed on the inclusive DIS data and the updated 
$c$-quark HERA data are in agreement with such an 
enhancement that demonstrates a consistent treatment of the DIS data 
in the present analysis. Finally, 
a good description of the non-resonant 
DY $\gamma^*/Z$-production data, which  are included into our analysis 
for the first time, is achieved provided photon-initiated lepton pair 
production is taken into account.

%
%

\end{document}